\documentclass[10pt,aps,prd,twocolumn,showpacs,superscriptaddress,eqsecnumnofootinbib,eqsecnum,longbibliography,nofootinbib]{revtex4-1}
\usepackage{mathrsfs,amsmath,amsthm,latexsym,amssymb,amsfonts,epsfig,txfonts,cancel}
\usepackage{xcolor}
\definecolor{navy}{RGB}{0,0,150}
\usepackage[colorlinks, linkcolor=cyan, citecolor=navy, urlcolor=navy, plainpages=false, pdfstartview=FitH]{hyperref}
\makeatletter
\newcommand{\doitext}{}

\renewcommand*{\doi}{%
  \begingroup 
  \lccode`\~=`\#\relax 
  \lowercase{\def~{\#}}%
  \lccode`\~=`\_\relax
  \lowercase{\def~{\_}}%
  \lccode`\~=`\<\relax 
  \lowercase{\def~{\textless}}%
  \lccode`\~=`\>\relax 
  \lowercase{\def~{\textgreater}}%
  \lccode`\~=0\relax 
  \catcode`\#=\active 
  \catcode`\_=\active 
  \catcode`\<=\active 
  \catcode`\>=\active 
  \@doi
}

\def\@doi#1{%
  \let\#\relax
  \let\_\relax
  \let\textless\relax 
  \let\textgreater\relax 
  \edef\x{\toks0={{#1}}}%
  \x
  \edef\#{\@percentchar23}%
  \edef\_{_}%
  \edef\textless{\@percentchar3C}
  \edef\textgreater{\@percentchar3E}
  \edef\x{\toks1={\noexpand\href{http://dx.doi.org/#1}}}%
  \x
  \edef\x{\endgroup\the\toks1 {\doitext\the\toks0}}%
  \x
}
\makeatother
\usepackage{appendix}
\allowdisplaybreaks
\newcommand{\GZU}{School of Physics, Guizhou University, Guiyang 550025, China}
\newcommand{\BNU}{Department of Physics, Beijing Normal University, Beijing 100875, China}

\begin{document}

\title{Loop quantum cosmology from an alternative Hamiltonian}

\author{Jinsong Yang}
\email{jsyang@gzu.edu.cn}
\affiliation{\GZU}

\author{Cong Zhang}
\email{zhang.cong@mail.bnu.edu.cn}
\affiliation{\BNU}

\author{Yongge Ma}
\thanks{Corresponding author}
\email{mayg@bnu.edu.cn}
\affiliation{\BNU}


\begin{abstract}
In this paper, a new Hamiltonian constraint operator for loop quantum cosmology is constructed by using the Chern-Simons action. The quantum dynamics of the $k=0$ cosmological model with respect to a massless scalar field as an emergent time is determined by a difference equation. The semiclassical analysis shows that the new quantum dynamics has the correct classical limit, and the classical big-bang singularity is again replaced by the quantum bounce. Interestingly, the inverse time evolution of the flat Friedmann-Robertson-Walker cosmological model determined by the effective Hamiltonian will be bounced to an asymptotic de Sitter universe.

\vspace{0.3cm}\noindent
DOI:~\doi{10.1103/PhysRevD.100.064026}
\end{abstract}


\maketitle

\section{Introduction}

A key motivation of studying quantum gravity is to cure the classical singularities predicted by general relativity (GR), by combining the ideas of GR and quantum theory in a consistent way. As a nonperturbative approach to quantum gravity, loop quantum gravity (LQG) has made remarkable progress in the past thirty years (see \cite{Rovelli:2004tv,Thiemann:2007pyv} for books, and \cite{Thiemann:2002nj,Ashtekar:2004eh,Han:2005km,Giesel:2012ws} for articles). According to LQG, a spacetime consists of fundamental units of spacetime quanta and the spectra of the geometrical operators corresponding to the classical length, area, and volume turn out to be discrete \cite{Rovelli:1994ge,Ashtekar:1996eg,Ashtekar:1997fb,Thiemann:1996at,Ma:2010fy}. In spite of these achievements, the dynamics of LQG is still an open issue. There are some interesting attempts to quantize the Hamiltonian constraint in the canonical approach \cite{Thiemann:1996aw,Thiemann:1997rt,Yang:2015zda,Alesci:2015wla}, and some properties of the resulted operators are studied \cite{Alesci:2011ia,Thiemann:2013lka,Zhang:2018wbc,Zhang:2019dgi}. Beyond these proposals, the idea to use Chern-Simons action to construct the Hamiltonian constraint is proposed in \cite{Soo:2005gw,Soo:2007hj}, which deserves further investigating. 
 
The ideas and techniques in the full theory of LQG have been successfully applied to the cosmological models of loop quantum cosmology (LQC) \cite{Ashtekar:2003hd}. Much progress, including the notable classical big-bang singularity resolution \cite{Bojowald:2001xe,Ashtekar:2006rx,Ashtekar:2006uz,Ashtekar:2006wn,Ding:2008tq,Yang:2009fp,Assanioussi:2018hee}, has been made in this area. The issue of big-bang singularity resolution in LQC was first discussed in \cite{Bojowald:2001xe,Ashtekar:2003hd} by noticing that the quantum Einstein equation is still well defined when the evolution approaches to the classical big-bang region, and the quantum evolution can ``evolve'' right through the singularity \cite{Bojowald:2001xe,Ashtekar:2003hd}. Later on, by the analytical and numerical methods developed in \cite{Ashtekar:2006rx,Ashtekar:2006uz,Ashtekar:2006wn,Ding:2008tq}, the dynamics at the quantum level was studied more profoundly, which shows that the classical big bang is replaced by a quantum bounce in LQC model of gravity coupled to a massless scalar field.
 
In LQG, the Hamiltonian constraint is usually decomposed as a summation of two terms: the Euclidean term and the Lorentzian term. These two terms were first regularized and quantized as operators in \cite{Thiemann:1996aw,Thiemann:1997rt} by using the Thiemann's trick, which was adapted afterward to define the Hamiltonian operator in LQC \cite{Ashtekar:2006rx,Ashtekar:2006uz,Ashtekar:2006wn,Yang:2009fp}. It is noticed that in \cite{Ashtekar:2006rx,Ashtekar:2006uz,Ashtekar:2006wn} one only considered the Thiemann's trick for the Euclidean term since the Lorentzian term is classically proportional to the Euclidean one in the spatially flat and homogeneous cosmological models. However, taking account of the fact that the Lorentzian term was quantized in a rather different way than that of the Euclidean term in full LQG, to inherit more features from the full theory, the Lorentzian term was treated independently in the cosmological models proposed in \cite{Yang:2009fp} by using the Thiemann's trick as in LQG. Recently, it is shown that the effective Hamiltonian of one of the models proposed in \cite{Yang:2009fp} can be reproduced by a suitable semiclassical analysis of Thiemann's Hamiltonian in full LQG \cite{Dapor:2017rwv}. A notable difference between the treatment without the Lorentzian term and the treatment with the Lorentzian term is that the former one leads to a symmetric bounce scenario while the later one an asymmetric bounce which relates the flat Friedmann-Robertson-Walker (FRW) cosmological model with an asymptotic de Sitter universe \cite{Assanioussi:2018hee,Li:2018opr}.

In this paper, we consider the idea to use Chern-Simons action to construct the Euclidean Hamiltonian constraint operator for LQG and apply this treatment to the LQC model. An interesting question is what kind of evolution can be obtained by using the new Hamiltonian constraint operator. We will show that the new proposed Hamiltonian constraint operator can also drive an asymmetric quantum bounce evolution, and the retrieving evolution of the flat FRW cosmological model will be bounced to a de Sitter universe asymptoticly. 

\section{Classical dynamics}

\subsection{Hamiltonian constraint from Chern-Simons action in LQG}

In the Hamiltonian formulation of GR, the spacetime manifold $M$ is splitted as $M=\varmathbb{R}\times\Sigma$ where $\Sigma$ denotes a three-dimensional spacelike manifold with arbitrary topology. The classical phase space consists of the Ashtekar-Barbero variables $(A^i_a(x),\tilde{E}^a_i(x))$, where $A^i_a(x)$ is a $SU(2)$-connection and $\tilde{E}^a_i(x)$ is a $\mathfrak{su}$(2)-valued vector density field of weight 1. Here $a,b,c...=1,2,3$ are used to represent the spatial indices and $i,j,k...=1,2,3$, the internal $\mathfrak{su}(2)$-indices. The only nontrivial Poisson bracket is given by
\begin{align}\label{eqn:full-poisson}
\{A^i_a(x),\tilde{E}^b_j(y)\}=\kappa\gamma\delta^a_b\delta^i_j\delta^3(x,y)\,,
\end{align}
where $\kappa=8\pi{G}$ with $G$ being Newtonian's gravitational constant, and $\gamma$ is the Barbero-Immirzi parameter. The classical dynamics of GR is thus obtained by imposing constraints on this phase space, including the Gausssian constraint, the diffeomorphism constraint, and the Hamiltonian constraint. Since the Gaussian constraint and the diffeomorphism constraint are identically satisfied in homogeneous models of cosmology, we only need to consider the Hamiltonian constraint here. The Hamiltonian constraint of gravity can be written as
\begin{align}\label{eqn:full-hamilton}
{\cal H }_{\rm grav}&=\int_\Sigma{\rm d}^3x\, N \frac{\tilde{E}^{aj}\tilde{E}^{bk
}}{2\kappa\sqrt{\textrm{det}(q)}}\left[\epsilon_{ijk}F^{i}_{ab}
-2(1+\gamma)K^j_{[a}K^k_{b]}\right]\,,
\end{align}
where $F^i_{ab}$ is the curvature of connection $A^i_a$, $K^i_a$ represents the extrinsic curvature of the spatial manifold $\Sigma$ embedded in $M$, and $N$ is an arbitrary smearing function. Since the homogeneous space is considered in the current paper, the smearing function can be fixed to be $N=1$. The first term in Eq. \eqref{eqn:full-hamilton} takes the same form as that in the Euclidean theory, and hence it is called the Euclidean term of the Hamiltonian constraint and denoted as
\begin{align}\label{eu-hamilton}
 {\cal H}^{\rm E}=\int_\Sigma{\rm d}^3x\frac{\tilde{E}^{aj}\tilde{E}^{bk}}{2\kappa\sqrt{\textrm{det}(q)}}\epsilon_{ijk}F^{i}_{ab}\,.
\end{align}

In LQG, there is no operator corresponding to $A_a^i(x)$ itself. For passage to quantum theory, one has to express the classical Euclidean Hamiltonian constraint in terms of the elementary variables: the holonomy $h_e(A)$ of $A_a^i(x)$ along some curve $e$ and the flux $\tilde{E}_i(S)$ of $\tilde{E}^a_i(x)$ across some surface $S$, which have unambiguous quantum analogs. In the standard LQC, one used the Thiemann's trick to achieve the end, where the Euclidean part can be rewritten as
\begin{align}\label{eqn:A-V-poss}
{\cal H }^{\rm E}=-\frac{2}{\kappa^2\gamma}\int_\Sigma{\rm d}^3x\,{\cal S}\, \tilde{\epsilon}^{abc}\textrm{tr}(F_{ab}(x)\{A_c(x),V\})\,,
\end{align}
where ${\cal S}\equiv{\rm sgn}\left[\det(e^i_a)\right]$, $\tilde{\epsilon}^{abc}$ denotes the Levi-Civita symbol, and $V$ is the volume of an arbitrary bounded open region in $\Sigma$ containing the point $x$. However, in the current paper, we will choose another approach to express ${\cal H }^{\rm E}$ in terms of the Chern-Simons action defined on $\Sigma$, which reads
\begin{align}\label{eqn:C-S-action}
S_{\rm cs}=2\int_\Sigma{\textrm{tr}\left(\textbf{A}\wedge{{\rm d}\textbf{A}}+\frac{2}{3}\textbf{A}\wedge\textbf{A}\wedge\textbf{A}\right)}\,,
\end{align}
where $\textbf{A}$ is used to abbreviate the connection $A_a^i(x)\tau_i{\rm d}x^a$, here $\tau_i=-\frac{\rm i}{2}\sigma_i$ (with $\sigma_i$ being the Pauli matrices), and ${\rm d}\textbf{A}$ is the exterior differentiation of $\textbf{A}$. In order to obtain ${\cal H }^{\rm E}$, we define the Chern-Simons functional on the phase space as
\begin{align}\label{csaction}
 S_{\rm cs}(A,\tilde{E})&:=2\int_\Sigma{\cal S}\,{\rm tr}\left(\textbf{A}\wedge{{\rm d}\textbf{A}}+\frac{2}{3}\textbf{A}\wedge\textbf{A}\wedge\textbf{A}\right)\notag\\
&=-\frac{1}{2}\int_\Sigma {\rm d}^3x\, {\cal S}\,\tilde{\epsilon}^{abc}\left(F^i_{ab}A^i_c-\frac{1}{3}\epsilon_{ijk}A^i_aA^j_bA^k_c\right)\,.
\end{align}
Suppose that $\Sigma$ is compact. Following \cite{Soo:2005gw,Soo:2007hj}, a straightforward calculation gives the Poisson bracket between Chern-Simons functional $S_{\rm cs}$ and the volume $V_\Sigma$ of $\Sigma$ as
\begin{align}
\{S_{\rm cs}(A,\tilde{E}),V_\Sigma\}=2\int_\Sigma {\rm d}^3x\,{\cal S}\,\tilde{\epsilon}^{abc}\textrm{tr}\left(F_{ab}(x)\left\{A_c(x),V_\Sigma\right\}\right)\,.
\end{align}
Therefore, the Euclidean part of the Hamiltonian constraint \eqref{eqn:A-V-poss} can be re-expressed as
\begin{align}\label{eqn:cs-H(1)}
{\cal H }^{\rm E}=-\frac{1}{\kappa^2\gamma}\left\{S_{\rm cs}(A,\tilde{E}),V_\Sigma\right\}\,.
\end{align}

Finally, to write the complete constraint, we also need the matter part of the constraint. In the present work, a massless scalar field $T$ minimal coupled to gravity is considered. Using $p_T$ to denote the momentum conjugate to $T$, 
we have the Hamiltonian of the scalar field
\begin{align}
{\cal H }_{\rm M}&=\frac{1}{2}\int_\Sigma {\rm d}^3 x\left( \frac{p_T^2}{\sqrt{q}}+(\nabla T)^2\right).
\end{align}

\subsection{Classical dynamics in the flat FRW cosmological model}
\label{sec:class-isotropic}

Let us consider the spatially flat FRW model of cosmology. As in the standard treatment in LQC, an elementary cubic cell $\mathcal{V}$ has to be introduced to avoid the noncompact problem of the spatial manifold. Fix a fiducial metric ${}^o\!q_{ab}$ and denote by $V_{o}$ the volume of this elementary cell in this geometry. Thus, the canonical variables are reduced to \cite{Ashtekar:2003hd}
\begin{align}\label{eqn:c-p}
A^i_a=c V_o^{-\frac{1}{3}}\ {}^o\!\omega^i_a\,, \quad \tilde{E}^a_i=p V_o^{-\frac{2}{3}}\sqrt{{}^o\!q}\ {}^o\!e^a_i\,,
\end{align}
where $({}^o\!\omega^i_a,{}^o\!e^a_i)$ are a set of orthonormal cotriads and triads compatible with ${}^o\!q_{ab}$ and adapted to the edges of the elementary cell. Thus, the Poisson bracket of basic variables reads
\begin{align}\label{eqn:c-p-bracket}
\{c,p\}=\frac{\kappa\gamma}{3}\,,
\end{align}
where $c$ is related to the time derivative of scale factor, and $p$ is the physical area of a face of the elementary cell. Then in terms of the reduced variables, the Euclidean part of the Hamiltonian constraint \eqref{eu-hamilton} and the Chern-Simons funtional \eqref{csaction} become
\begin{align}\label{eu-cp}
{\cal H }^{\rm E}=\frac{3}{\kappa}c^2\sqrt{|p|}\,,
\end{align}
and
\begin{align}\label{cs-cp}
S_{\rm cs}(c,p)=-2{\rm sgn}(p)\,c^3\,.
\end{align} 
Thus, the Hamiltonian constraint in Eq. \eqref{eqn:full-hamilton} can be simplified as
\begin{align}
{\cal H }_{\rm grav}=-\frac{1}{\gamma^2}{\cal H }^{\rm E}=\frac{1}{\kappa^2\gamma^3}\{S_{\rm cs}(c,p),V\}\,,
\end{align}
where $V:={|p|}^{3/2}$ is the physical volume of the elemental cell ${\cal V}$.

The total Hamiltonian constraint, combining the gravity part with the matter part, is
\begin{align}\label{tot-hamilton}
{\cal H}_{\rm tot}=\frac{1}{\kappa^2\gamma^3}\left\{S_{\rm cs}(c,p),V\right\}+\int_{\cal V} {\rm d}^3x\sqrt{q}\,\rho_{\rm M}\,,
\end{align}
which is actually
\begin{align}\label{tot-cp}
{\cal H}_{\rm tot}=-\frac{3}{\kappa\gamma^2}c^2\sqrt{|p|}+\frac{p_T^2}{2|p|^{3/2}}\,.
\end{align}
Then the Hubble parameter can be calculated as
\begin{align}\label{hub}
H=\frac{\{p,{\cal H}_{\rm tot}\}}{2p}=\frac{1}{\gamma}\frac{c}{\sqrt{|p|}}{\rm sgn}(p)\,.
\end{align}
The Friedmann equation is then obtained by vanishing the Hamiltonian constraint, which is 
\begin{align}
H^2=\frac{\kappa}{3}\frac{p_T^2}{V^2}=:\frac{\kappa}{3}\,\rho_{\rm M}\,.
\end{align}

\section{New quantum dynamics in LQC}\label{kin}

\subsection{Kinematics of LQC}

The kinematical Hilbert space corresponding to the degrees of freedom of gravity is 
\begin{align}
 {\mathscr H}_{\rm grav}=L^2(\varmathbb{R}_{\mathrm{Bohr}},{\mathrm d}\mu_{\mathrm{Bohr}})\,,
\end{align}
where $\varmathbb{R}_{\rm Bohr}$ is the Bohr compactification of $\varmathbb{R}$ and ${\mathrm d}\mu_{\mathrm{Bohr}}$ is the Haar measure. 
There are two fundamental operators in this Hilbert space: $\hat{p}$ which represents the area of each side of the elementary cell and ${\widehat{\exp{({\rm i}\lambda c)}}}$ which is the building block to reconstruct the holonomy of the reduced connection $A_a^i(x)$ along an edge parallel to the triad ${}^o\!e^a_i$. Following the improved scheme in \cite{Ashtekar:2006wn}, it is useful to introduce a new operator
\begin{align}\label{v-operator-def}
\hat{v}=\frac{\textrm{sgn}(\hat{p})|\hat{p}|^{3/2}}{2\pi\gamma\ell^2_{\rm p}\sqrt\Delta}\,,
\end{align}
where $\ell_{\rm p}\equiv\sqrt{G\hbar}$ is the Planck length and $\Delta\equiv 2\sqrt{3}\pi\gamma\ell^2_{\rm p}$ denotes the area gap in full LQG. $\hat{v}$ is a dimensionless variable representing the physical volume of the elementary cell. We will work with the representation where the operator $\hat{v}$ is diagonalized. Eigenstates of $\hat{v}$, denoted as $|v\rangle$, are labeled by real numbers $v$. The orthonormal relation among these eigenstates is given by
\begin{align}
\langle v|v'\rangle=\delta_{v,v'}\,,
\end{align}
where $\delta_{v,v'}$ is the Kronecker delta. Thus, a general sate in ${\mathscr H}_{\rm grav}$ is expressed as a countable sum: $|\psi\rangle=\sum\psi_n|v_n\rangle$, and the inner product is
\begin{align}
\langle\psi^{(1)}|\psi^{(2)}\rangle=\sum\overline{\psi^{(1)}_n}\psi^{(2)}_n\,.
\end{align}

There are following two other useful operators in the kinematical Hilbert space ${\mathscr H}_{\rm grav}$. 
\begin{itemize}
\item[(i)] The first one is $\widehat{e^{{\rm i}b}}$, where 
\begin{align}
b:=\frac{\bar\mu{c}}{2}\,,
\end{align}
with $\bar\mu=\sqrt{\Delta/|p|}$. $\widehat{e^{{\rm i}b}}$ is the building block to formulate the holonomy $h^{(\bar\mu)}_i$ of the reduced connection $A_a^i(x)$ along an edge parallel to the triad ${}^o\!e^a_i$ whose length with respect to the physical metric is $\sqrt{\Delta} $. This means that the edge underlying the holonomy $h^{(\bar\mu)}_i$ takes the minimal length of the quantum geometry. Because of
\begin{align}
\{b,v\}=\frac{1}{\hbar}\,,
\end{align}
one has
\begin{align}
\widehat{e^{{\rm i}b}}\,|v\rangle=|v+1\rangle\,.
\end{align}
$\widehat{e^{{\rm i}b}}$ is related to the holonomy $\widehat{h}_i^{\bar\mu}$ according to the classical relation
\begin{align}\label{eqn:i-holonomy}
&h^{(\bar\mu)}_i=\cos(b)\,\mathbb{I}+2\sin(b)\,\tau_i\,.
\end{align}

\item[(ii)] The other one is the operator $\hat V$ representing the volume of the elementary cell. It reads from Eq. \eqref{v-operator-def}
\begin{align}
\hat{V}=2\pi\gamma\ell_{\rm p}^2\sqrt{\Delta}\,|\hat v|\,.
\end{align}

\end{itemize}

For the degrees of freedom of the scalar filed, we use the Sch\"ordinger quantization, where $\hat{T}$ is quantized to be a multiplication operator and $\hat{p}_T$ is the derivative operator, namely
\begin{align}
\hat{T}\psi(T)&=T\psi(T)\,,\qquad
\hat{p}_T\Psi(T)=-i\hbar\frac{{\mathrm d}}{{\mathrm d} T}\Psi(T)\,,
\end{align}
where $\Psi(T)\in L^2(\varmathbb{R},{\mathrm d} T)$. Then the total kinematical Hilbert space, combining those for gravity and the scalar field, is ${\mathscr H}_{\rm tot}={\mathscr H}_{\rm grav}\otimes L^2(\varmathbb{R},{\mathrm d} T)$.

\subsection{Hamiltonian constraint operator for gravity in LQC}

To construct the Hamiltonian constraint operator, we first express the Chern-Simons functional \eqref{csaction} in terms of holonomies as
\begin{align}\label{eqn:cs-discrete1}
S_{\rm cs}&=\frac{{\rm sgn}(p)}{\bar{\mu}^3}\left[\epsilon^{ijk}\textrm{tr}\left(h^{(\bar{\mu})}_{\square_{ij}}
\frac{h^{(\bar{\mu})}_k-h^{(\bar{\mu})^{-1}}_k }{2}\right)\right.\notag\\
&\hspace{1cm}\left.-
\frac{2}{3}\epsilon^{ijk}\textrm{tr}\left(\frac{h^{(\bar{\mu})}_i-h^{(\bar{\mu})^{-1}}_i}{2}\frac{h^{(\bar{\mu})}_j-h^{(\bar{\mu})^{-1}}_j }{2}\frac{h^{(\bar{\mu})}_k-h^{(\bar{\mu})^{-1}}_k }{2}\right)\right]\,,
\end{align}
where $h^{(\bar{\mu})}_{\square_{ij}}:=h^{(\bar{\mu})}_ih^{(\bar{\mu})}_jh^{(\bar{\mu})^{-1}}_ih^{(\bar{\mu})^{-1}}_j$ is the holonomy around the square ${\square_{ij}}$ in the $i$-$j$ plane spanned by a face of the elementary cell with $\bar{\mu}V_0^{1/3}$ being the length of its sides measured by fiducial metric ${}^oq_{ab}$. It should be noticed that in Eq. \eqref{eqn:cs-discrete1} we have taken into account the existence of the area gap \cite{Ashtekar:2006wn}. Actually, the right-hand side of Eq. \eqref{eqn:cs-discrete1} recovers exactly the classical expression of $S_{\rm cs}$ when the area of the squares ${\square_{ij}}$ goes to 0.

By substituting Eq. \eqref{eqn:i-holonomy} into Eq. \eqref{eqn:cs-discrete1}, we obtain
\begin{align}\label{eqn:cs-discrete2}
S_{\rm cs}=\frac{{\rm sgn}(p)}{\bar{\mu}^3}\left[8\sin^3(b)-6\sin^2{(2b)}\sin(b)\right]\,,
\end{align}
which leads to
\begin{align}\label{hamilton}
{\cal H }_{\rm grav}=\frac{1}{\kappa^2\gamma^3}\left\{\frac{{\rm sgn}(p)}{\bar{\mu}^3}
\left[8\sin^2(b)-6\sin^2(2b)\right]\sin(b),V\right\}\,.
\end{align}
According to Eq. \eqref{hamilton}, ${\cal H }_{\rm grav}$ involves not only $\sin^2(b)$ but also $\sin^2(2b)$. This result shares a similar feature with the alternative Hamiltonian constraint proposed in \cite{Yang:2009fp}, that it involves simultaneously both $\sin^2(b)$ and $\sin^2(2b)$ terms. The simultaneous existence of these two terms in the alternative Hamiltonian leads to an asymmetric bounce in the evolution of the Universe \cite{Dapor:2017rwv,Assanioussi:2018hee,Li:2018opr}, while the Hamiltonian in \cite{Ashtekar:2006rx} which contains only the $\sin^2(b)$ term results in a symmetric bounce. Therefore, it is expected that the quantum Hamiltonian constraint obtained from Eq. \eqref{hamilton} will result in an asymmetry bounce finally. The corresponding Hamiltonian constraint operator is obtained by replacing the classical variables with their quantum analogs, which reads
\begin{align}\label{eqn:H(1)-operator}
\hat{\cal H}_{\rm grav}=-\frac{{\rm i}\hbar}{32\gamma\sqrt{\Delta}}\,\hat{|v|}^{1/2}
\left[\hat{\cal B}\,\widehat{{\rm sgn}(v)}+\widehat{{\rm sgn}(v)}\,\hat{\cal B},\hat{|v|}\right]\hat{|v|}^{1/2}\,,
\end{align}
where
\begin{align}
\hat{\cal B}\equiv 8\widehat{\sin^3(b)}-6\widehat{\sin(2b)}\;\widehat{\sin(b)}\;\widehat{\sin(2b)}\,.
\end{align}
Its action on the state $|v\rangle$ is given by
\begin{align}\label{hamilton_gr_op}
\hat{\cal H}_{\rm grav}|v\rangle=&3f^+_5(v)|v+5\rangle
-7f^+_3(v)|v+3\rangle+6f^+_1(v)|v+1\rangle\notag\\
&+6f_1^-(v)|v-1\rangle
-7f^-_3(v)|v-3\rangle+3f^-_5(v)|v-5\rangle\,
\end{align}
with
\begin{align}
f_k^{\pm}(v)&=\pm\frac{\hbar}{128\gamma\sqrt{\Delta}}\left[{\rm sgn}(v\pm k)+{\rm sgn}(v)\right]\notag\\
&\hspace{2cm}\times(|v\pm k|-|v|)\sqrt{|v(v\pm k)|}\,.
\end{align}

\subsection{The total Hamiltonian operator}

According to Eq. \eqref{tot-cp}, the Hamiltonian for the sector of the scalar field is given by
\begin{align}\label{eqn:matter-hamilton}
{\cal H }_{\rm M}=\frac{p^2_T}{2V}\,.
\end{align}
As in \cite{Ashtekar:2006wn}, the inverse volume operator corresponding to $1/V$ is given by 
\begin{align}\label{eq:inverseV}
\widehat{V^{-1}}\psi(v)=\frac{ B(v)}{2\pi\gamma\sqrt{\Delta}\ell_{\mathrm p}^2}\psi(v)\,,
\end{align}
where
\begin{align}
B(v)\equiv \left(\frac{3}{2}\right)^3|v|
\left|{|v+1|^{{1}/{3}}}-{|v-1|^{{1}/{3}}}\right|^3\,.
\end{align}
Then by recalling that the Hilbert space for the scalar field is given by the Sch\"ordinger quantization, the action of the Hamiltonian operator for the matter part reads
\begin{align}\label{hamilton_matter_op}
\hat{\cal H}_{\rm M}\psi(v;T)=-\frac{\hbar^2}{4\pi\gamma\sqrt{\Delta}\ell_{\mathrm p}^2}B(v)\partial^2_T\psi(v;T)\,.
\end{align}
Finally, by using Eqs. \eqref{hamilton_gr_op} and \eqref{hamilton_matter_op}, we have the total Hamiltonian constraint equation
\begin{align}
 \hat{\cal H}_{\rm tot}\psi(\nu;T)\equiv\left(\hat{\cal H}_{\rm grav}+\hat{\cal H}_{\rm M}\right)\psi(\nu;T)=0\,,
\end{align}
which gives the dynamics of the current system as
\begin{align}\label{eq:dynmceqt}
\partial^2_T\psi(v;T)=&[B(v)]^{-1}\big[+6C_1^+(v)\psi(v+1;T)+6C_1^-(v)\psi(v-1;T)\notag\\
&\hspace{1cm}-7C_3^+(v)\psi(v+3;T)-7C_3^-(v)\psi(v-3;T)\notag\\
&\hspace{1cm}+3C_5^+(v)\psi(v+5;T)+3C_5^-\psi(v-5;T)\big]\notag\\
=:&\widehat{\Theta} \psi(v;T)\,,
\end{align}
where
\begin{align}\label{Ck:def}
C_k^{\pm}(v)=\pm\frac{\pi G}{32}\left[{\rm sgn}(v\pm k)+{\rm sgn}(v)\right](|v\pm k|-|v|)\sqrt{|v(v\pm k)|}\,.
\end{align}

\section{Semiclassical analysis of the quantum dynamics}

We will choose a coherent state peaked at some point $(b_0,v_0,T_0,p_T)$ in the classical phase space to calculate the expectation values of the quantum constraint. The coherent state is chosen to be 
\begin{align}\label{eqn:dual-semiclassical-state}
\left(\Psi_{(b_o,v_o,T_0,p_T)}\right|&:=\int {\rm d}T\sum_{v\in\varmathbb{R}}e^{-\frac{\epsilon^2}{2}(v-v_o)^2} e^{{\rm i}b_o(v-v_o)} e^{-\frac{\sigma^2}{2}(T-T_0)^2}\notag\\
&\hspace{2cm}\times e^{\frac{\rm i}{\hbar}p_T(T-T_0)}(v|\otimes (T|\,,
\end{align}
where $\epsilon$ and $\sigma$ are the Gaussian spreads in the gravitational sector and scalar field sector, respectively. It is noticed that this state lives in the algebraic dual space of some dense set in the kinematical Hilbert space. For practical calculations, we only need to use the shadow of the state $\left|\Psi_{(b_o,v_o,T_0,p_T)}\right\rangle$ on the regular lattice with spacing 1 as
\begin{align}\label{eqn:shadow-grav}
\left|\Psi_{(b_o,v_o,T_0,p_T)}\right\rangle&:=\int{\rm d}T\sum_{n\in\mathbb{Z}}e^{-\frac{\epsilon^2}{2}(n-v_o)^2}e^{-{\rm i}b_o(n-v_o)} e^{-\frac{\sigma^2}{2}(T-T_0)^2}\notag\\
&\hspace{2cm}\times e^{-\frac{\rm i}{\hbar}p_T(T-T_0)}|n\rangle\otimes |T\rangle\,.
\end{align}
As discussion in \citep{Taveras:2008ke}, the parameters $v_o$ and $b_o$ in the coherent state \eqref{eqn:dual-semiclassical-state} are restricted to satisfying $v_o\gg 1$ and $b_o\ll 1$ so that the corresponding volume $V_o=2\pi\gamma \ell_{\rm P}^2\sqrt{\Delta}v_o\gg \ell_{\rm P}^3$ and expansion velocity $\dot{a}\ll 1$. Moreover, restrictions that $v_o\epsilon\gg 1$, $\epsilon\ll b_o$, $\phi\gg \sigma$ and $p_T\sigma\gg 1$ are also required so that the state is sharply peaked.

Although there is no operator corresponding to $b$ in loop quantization, one may define an approximation $\hat{b}:=(\widehat{e^{{\rm i}b}}-\widehat{e^{-{\rm i}b}})/(2{\rm i})$, which agrees approximately with the classical $b$ when $b\ll 1$. Using the shadow state scheme \cite{Ashtekar:2002sn}, the expectation values in the state $(\Psi|$ are calculated as
\begin{align}
\begin{aligned}
\langle b\rangle =e^{-\frac{1}{4}\epsilon^2}\sin{b_o}\,, \;\;\langle v\rangle=v_o\,, \;\;\langle T\rangle=T\,, \;\;\langle p_T\rangle=p_T\,.
\end{aligned}
\end{align}

Now let us calculate the expectation value of $\hat{\cal H}_{\rm grav}$. Without loss of generality, we drop the matter part in the shadow state \eqref{eqn:shadow-grav}. Then the action of the operator $\hat{\cal H}_{\rm grav}$ in Eq. \eqref{hamilton_gr_op} on the shadow state is given by
\begin{align}
\hat{\cal H}_{\rm grav}|\Psi\rangle&=\sum_{n\in\mathbb{Z}}e^{-\frac{\epsilon^2}{2}(n-v_o)^2}e^{-{\rm i}b_o(n-v_o)}\notag\\
&\quad\times\sum_{k\in\{1,3,5\}}\lambda_k\left( f_{k}^{+}(n)
|n+ k\rangle+f_k^-(n)|n-k\rangle\right)
\end{align}
with $\lambda_1=6,\,\lambda_3=-7,\,\lambda_5=3$. Note that $f_k^+(n)=f^-_k(n+k)$. We then get
\begin{align}
(\Psi|\hat{\cal H }_{\rm grav}|\Psi\rangle&=2\sum\limits_k\lambda_ke^{-\frac{\epsilon^2}{4}k^2}\cos{(kb_o)}\sum\limits_{n\in\mathbb{Z}}f^+_k(n)e^{-\epsilon^2\left(n-v_o+\frac{k}{2}\right)^2}\notag\\
&=:2\sum\limits_k\lambda_ke^{-\frac{\epsilon^2}{4}k^2}\cos{(kb_o)}F_k\,.
\end{align}
Applying the Poisson resummation formula, we obtain
\begin{align}
F_k&=\sum\limits_{n\in\mathbb{Z}}e^{-\frac{\pi^2n^2}{\epsilon^2}}e^{-{\rm i}2\pi n (v_o-\frac{k}{2})}\int_{-\infty}^\infty {\rm d}x f_k^+(x)e^{-\epsilon^2\left(x-v_o+\frac{k}{2}+\frac{{\rm i}n\pi}{\epsilon^2}\right)^2}\notag\\
&=I_k^0+O\left(e^{-\pi^2/\epsilon^2}\right)\,,
\end{align}
where
\begin{align}
I_k^0&=\int^\infty_{-\infty}{\rm d}xf_k^+(x-\frac{k}{2})e^{-\epsilon^2(x-v_o)^2}\\
&=\frac{\hbar v_o^3}{128\gamma\sqrt{\Delta}}\int_{-\infty}^\infty {\rm d}x e^{-(\epsilon v_o)^2(x-1)^2}\left[{\rm sgn}\left(x+\frac{k}{2v_o}\right)\right.\notag\\
&\left.+{\rm sgn}\left(x-\frac{k}{2v_o}\right)\left(\left|x+\frac{k}{2v_o}\right|-\left|x-\frac{k}{2v_o}\right|\right)\sqrt{\left|x^2-\frac{k^2}{4v_o^2}\right|}\right]\,.
\end{align}
By the steepest decent method, we get
\begin{align}
I_k^0=\frac{\hbar k v_o \sqrt{\pi}}{64\gamma\sqrt{\Delta}\,\epsilon}+O(1/v_o)+O\left(e^{-\epsilon^2v_o^2}\right)\,.
\end{align}
Thus, we obtain
\begin{align}
\langle \hat{\cal H }_{\rm grav}\rangle&=\frac{\langle \Psi|\hat{\cal H }_{\rm grav}|\Psi\rangle}{\langle \Psi|\Psi\rangle}\approx \frac{\hbar v_o}{32\gamma \sqrt{\Delta}}\sum\limits_kk\lambda_ke^{-\frac{\epsilon^2}{4}k^2}\cos(kb_o)\notag\\
&= \frac{\hbar v_o}{32\gamma\sqrt{\Delta}}
\left[
15e^{-\frac{25}{4}\epsilon^2}\cos{(5b_0)}
-21e^{-\frac{9}{4}\epsilon^2}\cos{(3b_o)}\right.\notag\\
&\left.\hspace{1.7cm}+6e^{-\frac{1}{4}\epsilon^2}\cos{(b_o)}
\right]\,.
\end{align}

Taking into account the result for the matter sector given in \citep{Ding:2008tq}, we finally have the expectation value for the total Hamiltonian constraint
\begin{align}\label{eq:PreefectiveHa}
\langle \hat{\cal H}_{\rm tot} \rangle\approx &\frac{\hbar v}{32\gamma\sqrt{\Delta}}
\left[
15e^{-\frac{25}{4}\epsilon^2}\cos{(5b)}
-21e^{-\frac{9}{4}\epsilon^2}\cos{(3b)}\right.\notag\\
&\hspace{4.2cm}\left.+6e^{-\frac{1}{4}\epsilon^2}\cos{(b)}\right]\notag\\
&+\frac{1}{4\pi\gamma\sqrt{\Delta}\,\ell_{\rm P}^2}\left(p_T^2+\frac{\hbar}{2\sigma^2}\right)\left[\frac{1}{v}+ O\left(v^{-3},v^{-3}\epsilon^{-2}\right)\right]\,,
\end{align}
where the subscript $o$ has been dropped.

\section{Dynamical analysis}

\begin{figure}
  \includegraphics[width=\columnwidth]{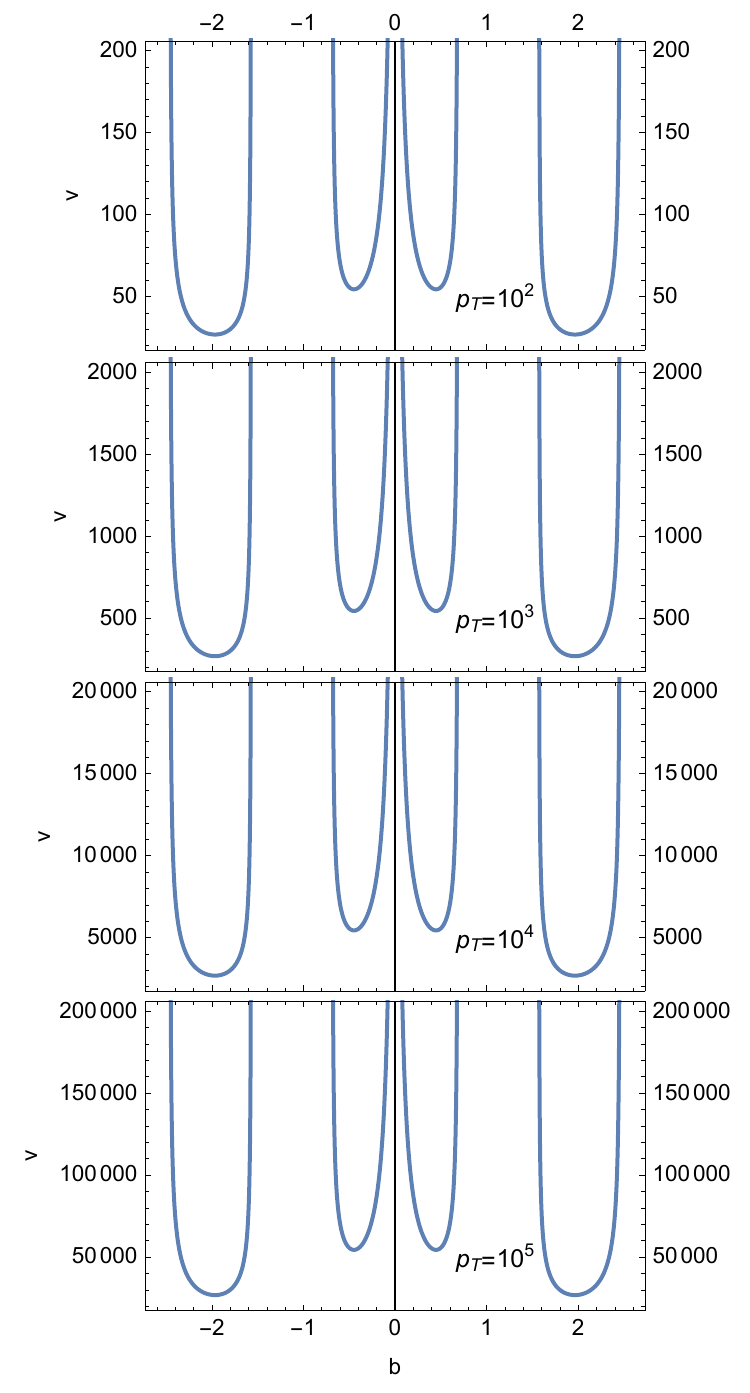}
  \caption{Plots of the constraint equation ${\cal H}^{\rm eff}_{\rm tot}(b,v,p_T)=0$ for different values of $p_T$ (the similar picture in quality holds for other values of $p_T$), where $G=\hbar=1$, $\gamma=0.2375$. It shows that $v=0$ cannot be a solution.}
  \label{figures:b-v}
\end{figure}

Consider the leading order of the effective Hamiltonian constraint given by Eq. \eqref{eq:PreefectiveHa}, which is
\begin{align}
\label{eff-Hamiltonian-eq}
{\cal H}^{\rm eff}_{\rm tot}\equiv\langle\hat{\cal H}_{\rm tot}\rangle&=-\frac{3\hbar}{8\gamma\sqrt{\Delta}}v\left[3\cos(b)+5\cos(3b)\right]\sin^2(b)\notag\\
&\quad+\frac{p_T^2}{4\pi\gamma\ell_{\rm P}^2\sqrt{\Delta}\, v}\,.
\end{align} 
It is easy to see that $p_T$ is a constant of motion. Thus, $T$ is a monotonic function of the cosmological time. Then the matter field $T$ can be regarded as an internal clock with respect which the relative evolution can be defined. In addition, by the form of the effective Hamiltonian constraint \eqref{eff-Hamiltonian-eq}, $v=0$ can never be a solution to the constraint equation ${\cal H}^{\rm eff}_{\rm tot}(b,v)=0$, which indicates that the big-bang singularity where $v=0$ will be resolved by the effective Hamiltonian constraint. For a given $p_T$, the equation ${\cal H}^{\rm eff}_{\rm tot}(b,v)=0$ is plotted in Fig. \ref{figures:b-v}, in which the conclusion that $v\neq 0$ is shown.

According to the effective Hamiltonian constraint, the matter density can be expressed as
\begin{align}\label{rho-relation}
 \rho_T(v)&=\frac{p_T^2}{2V^2}=\frac{p_T^2}{8\pi^2G^2\hbar^2\gamma^2\Delta\, v^2}\notag\\
 &\stackrel{{\cal H}^{\rm eff}_{\rm tot}=0}{=}\frac{3}{16\pi G\gamma^2\Delta}\left[3\cos(b)+5\cos(3b)\right]\sin^2(b)\notag\\
&\equiv\rho^{\rm eff}_T(b)\,,
\end{align}
which takes its critical matter density $\rho_{\rm crit}$, defined as the maximal value of $\rho^{\rm eff}_T(b)$ at $b=\arcsin\left(\frac{\sqrt{13-\sqrt{69}}}{5}\right)$, as
 \begin{align}
 \rho_{\rm crit}=\frac{3\sqrt{7092-759\sqrt{69}}}{625\pi G\gamma^2\Delta}\approx\frac{1}{8\pi G\gamma^2\Delta}\,.
\end{align}

Now let us study the asymptotic behavior of the effective dynamics in the classical region, namely the large $v$ region. For $v\rightarrow \infty$, the matter density $\rho^{\rm eff}_T(b)$ in Eq. \eqref{rho-relation} goes to zero, which leads to
\begin{align}\label{eqn:bb}
b\in\left(-\frac{\pi}{2},\frac{\pi}{2}\right)\;\rightarrow\;
  \left\{
  \begin{array}{ll}
  b^\pm_{\rm c,I}\equiv0^\pm\\
  b^\pm_{\rm c,II}\equiv\pm\arcsin\left(\sqrt{\frac25}\right)
 \end{array}\right.\,.
\end{align}
The above result shows that there are two types of classical universe, the type-I and the type-II universes. In the following, we focus on the positive solution corresponding to our universe and omit the symbol $+$. Moreover, Fig. \ref{figures:b-v} shows that the two universes are in fact connected. Expanding $\rho^{\rm eff}_T(b)$ at $b_{\rm c,I}$ and $b_{\rm c,II}$ up to the second order, one obtains the classical behavior of the matter density
\begin{align}
\rho^{\rm eff}_T\rightarrow
  \left\{
  \begin{array}{ll}
  \rho_{\rm c,I}\equiv\xi b^2\\
  \rho_{\rm c,II}\equiv-\frac{3\xi}{5\sqrt{5}}\left[2\sqrt{2}\left(b-b_{\rm c,II}\right)+3\sqrt{3}\left(b-b_{\rm c,II}\right)^2\right]
 \end{array}\right.\,,
\end{align}
where $\xi\equiv\frac{3}{2\pi G\gamma^2\Delta}$, and the classical behavior of the effective Hamiltonian constraint
\begin{align}
{\cal H}^{\rm eff}_{\rm tot}\rightarrow
  \left\{
  \begin{array}{ll}
  {\cal H}_{\rm tot, c,I}\equiv-\frac{3\hbar}{\gamma\sqrt{\Delta}}b^2v+\frac{p_T^2}{2V}\\
  {\cal H}_{\rm tot, c,II}\equiv\frac{9\hbar}{5\gamma\sqrt{5\Delta}}\left[2\sqrt{2}\left(b-b_{\rm c,II}\right)\right.\\
  \hspace{2.2cm}\left.+3\sqrt{3}\left(b-b_{\rm c,II}\right)^2\right]v+\frac{p_T^2}{2V}
 \end{array}\right.\,.
\end{align}
Substituting these asymptotic expressions into the Friedmann equation
\begin{align}
 H_{\rm eff}^2=\left(\frac{\dot{v}}{3v}\right)^2=\left(\frac{\left\{v,{\cal H}^{\rm eff}_{\rm tot}\right\}}{3v}\right)^2=\left(-\frac{1}{3v\hbar}\frac{\partial {\cal H}^{\rm eff}_{\rm tot}}{\partial b}\right)^2\,,
\end{align}
one can get the classical behavior of Hubble parameter
\begin{align}\label{H_eff2}
 H_{\rm eff}^2\rightarrow
 \left\{
 \begin{array}{ll}
  H^2_{\rm c,I}\equiv\frac{8\pi G}{3}\rho_T\\
  H^2_{\rm c,II}\equiv\frac{8\pi G}{3}\left[-\frac{9\sqrt{3}}{5\sqrt{5}}\,\rho_T+\rho_\Lambda\right]
 \end{array}
 \right.\,,
\end{align}
where 
\begin{align}
 \rho_\Lambda\equiv\frac{27}{125\pi G\gamma^2\Delta}\,.
\end{align}
Equation \eqref{H_eff2} implies that in the type-I region, which corresponds to the case $b\to 0$ in Eq. \eqref{eqn:bb}, the effective dynamics behaves as the standard FRW cosmology coupled to a scalar field, while in the type-II region, which corresponds to the other case in Eq. \eqref{eqn:bb}, the effective dynamics behaves as a FRW universe coupled to a scalar field with negative energy density and a positive cosmological constant
\begin{align}
 \Lambda=\frac{216}{125\gamma^2\Delta}\,.
\end{align}

\begin{figure}
  \includegraphics[width=\columnwidth]{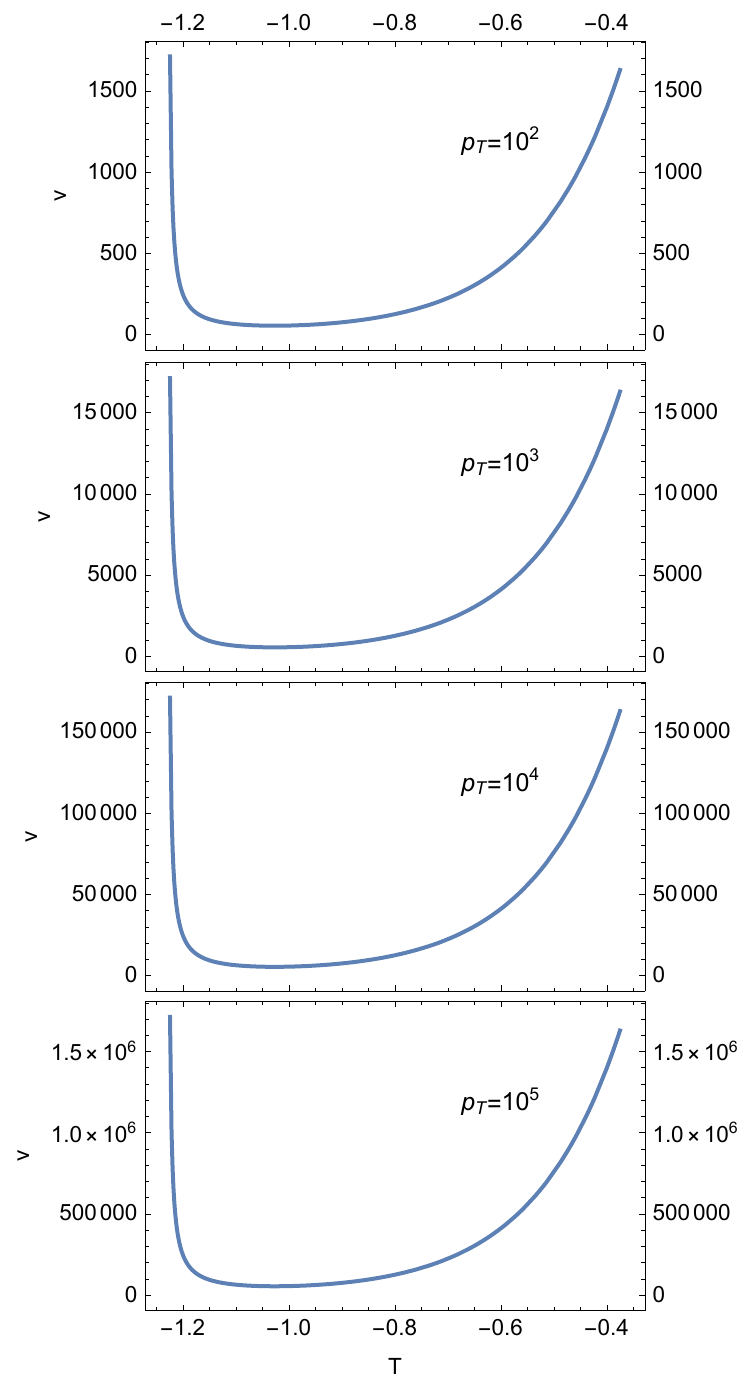}
  \caption{Plots of the relative evolution of $v$ with respect to the scalar field $T$ given by the effective Hamiltonian constraint for different values of $p_T$ (the similar picture in quality holds for other values of $p_T$), where $G=\hbar=1$, $\gamma=0.2375$.}
  \label{figures:phi-v}
\end{figure}

By using the Hamiltonian constraint equation and the dynamical equation,
\begin{align}
\label{dphi-dv}
 \frac{{\rm d}T}{{\rm d}v}=\frac{\left\{T,{\cal H}^{\rm eff}_{\rm tot}\right\}}{\left\{v,{\cal H}^{\rm eff}_{\rm tot}\right\}}=\frac{16p_T}{3\pi\hbar Gv^2\left[2\sin(b)-21\sin(3b)+25\sin(5b)\right]}\,,
\end{align}
the relative evolution of $v$ with respect to the scalar field $T$ can be calculated numerically. The solution is plotted in Fig. \ref{figures:phi-v}. It shows that the classical big-bang singularity is again replaced by a quantum bounce. Moreover, the time retrieving of the dynamical solutions shown in Fig. \ref{figures:phi-v} gives an evolution which bounces from the FRW universe to a de Sitter universe, i.e. an accelerating universe.

\section{Summary and discussion}

In the previous sections, the Chern-Simons action was employed to regularize the Euclidean Hamiltonian constraint of GR, and then a well-defined corresponding operator could be expected for LQG. To test this idea, the treatment was applied to the $k=0$ cosmological model with a massless scalar field, for which the Euclidean term of the gravitational Hamiltonian constraint is classically proportional to the Lorentzian term. Therefore, in our LQC model only the Euclidean term of the Hamiltonian is quantized for simplification. The quantum dynamics of the LQC model is determined by the difference equation \eqref{eq:dynmceqt}. It is shown by Eq. \eqref{eq:PreefectiveHa} that the quantum Hamiltonian has the correct classical limit by the semiclassical analysis. The effective dynamic gives a quantum bounce resolving the classical big-bang singularity as shown in Fig. \ref{figures:phi-v}.

Alternatively, one can also treat the Lorentzian part of the Hamiltonian constraint independently and add it into the Hamiltonian. This will lead to an alternative effective Hamiltonian constraint as follows:
\begin{align}
\label{Full-eff-Hamiltonian-eq}
{\cal H}^{\rm eff}_{\rm Full, tot}=&\frac{3\hbar\gamma}{8\sqrt{\Delta}}v\left[3\cos(b)+5\cos(3b)\right]\sin^2(b)\notag\\
&-\frac{3\hbar(1+\gamma^2)}{4096\sqrt{\Delta}\,\gamma}v\left[2\sin(b)-21\sin(3b)+25\sin(5b)\right]^2\notag\\
&+\frac{p_T^2}{4\pi\gamma\ell_{\rm P}^2\sqrt{\Delta}\,v}\,.
\end{align}
We leave the dynamics driven by the alternative Hamiltonian constraint \eqref{Full-eff-Hamiltonian-eq} for further study.

\begin{acknowledgments}
J. Y. would like to thank Professor Chopin Soo for useful discussions. This work is supported in part by NSFC Grants No. 11765006, No. 11875006, and No. 11961131013.
\end{acknowledgments}


%


\begin{thebibliography}{34}%

\bibitem{Rovelli:2004tv}
C.~Rovelli, \href{http://dx.doi.org/10.1017/CBO9780511755804}{{Quantum
  Gravity}}.
\newblock Cambridge University Press, Cambridge, United Kingdom, 2004.

\bibitem{Thiemann:2007pyv}
T.~Thiemann, \href{http://dx.doi.org/10.1017/CBO9780511755682}{{Modern
  Canonical Quantum General Relativity}}.
\newblock Cambridge University Press, Cambridge, United Kingdom, 2007.

\bibitem{Thiemann:2002nj}
T.~Thiemann, {Lectures on loop quantum gravity},
  \href{http://dx.doi.org/10.1007/978-3-540-45230-0_3}{Lect. Notes Phys.
  {\bfseries 631}, 41 (2003)}, \href{http://arxiv.org/abs/gr-qc/0210094}{{
  [arXiv:gr-qc/0210094}]}.

\bibitem{Ashtekar:2004eh}
A.~Ashtekar and J.~Lewandowski, {Background independent quantum gravity: a
  status report}, \href{http://dx.doi.org/10.1088/0264-9381/21/15/R01}{Class.
  Quant. Grav. {\bfseries 21}, R53 (2004)},
  \href{http://arxiv.org/abs/gr-qc/0404018}{{ [arXiv:gr-qc/0404018}]}.

\bibitem{Han:2005km}
M.~Han, Y.~Ma, and W.~Huang, {Fundamental structure of loop quantum gravity},
  \href{http://dx.doi.org/10.1142/S0218271807010894}{Int. J. Mod. Phys. D
  {\bfseries 16}, 1397 (2007)}, \href{http://arxiv.org/abs/gr-qc/0509064}{{
  [arXiv:gr-qc/0509064}]}.

\bibitem{Giesel:2012ws}
K.~Giesel and H.~Sahlmann, {From classical to quantum gravity: Introduction to
  loop quantum gravity}, \href{http://dx.doi.org/10.22323/1.140.0002}{Proc.
  Sci. {\bfseries QGQGS2011}, 002 (2011)},
  \href{http://arxiv.org/abs/1203.2733}{{ [arXiv:1203.2733}]}.

\bibitem{Rovelli:1994ge}
C.~Rovelli and L.~Smolin, {Discreteness of area and volume in quantum gravity},
  \href{http://dx.doi.org/10.1016/0550-3213(95)00150-Q}{Nucl. Phys. B
  {\bfseries 442}, 593 (1995)}, \href{http://arxiv.org/abs/gr-qc/9411005}{{
  [arXiv:gr-qc/9411005}]}.

\bibitem{Ashtekar:1996eg}
A.~Ashtekar and J.~Lewandowski, {Quantum theory of geometry: I. Area
  operators}, \href{http://dx.doi.org/10.1088/0264-9381/14/1A/006}{Class.
  Quant. Grav. {\bfseries 14}, A55 (1997)},
  \href{http://arxiv.org/abs/gr-qc/9602046}{{ [arXiv:gr-qc/9602046}]}.

\bibitem{Ashtekar:1997fb}
A.~Ashtekar and J.~Lewandowski, {Quantum theory of geometry II: Volume
  operators}, \href{http://dx.doi.org/10.4310/ATMP.1997.v1.n2.a8}{Adv. Theor.
  Math. Phys. {\bfseries 1}, 388 (1997)},
  \href{http://arxiv.org/abs/gr-qc/9711031}{{ [arXiv:gr-qc/9711031}]}.

\bibitem{Thiemann:1996at}
T.~Thiemann, {A length operator for canonical quantum gravity},
  \href{http://dx.doi.org/10.1063/1.532445}{J. Math. Phys. (N.Y.) {\bfseries
  39}, 3372 (1998)}, \href{http://arxiv.org/abs/gr-qc/9606092}{{
  [arXiv:gr-qc/9606092}]}.

\bibitem{Ma:2010fy}
Y.~Ma, C.~Soo, and J.~Yang, {New length operator for loop quantum gravity},
  \href{http://dx.doi.org/10.1103/PhysRevD.81.124026}{Phys. Rev. D {\bfseries
  81}, 124026 (2010)}, \href{http://arxiv.org/abs/1004.1063}{{
  [arXiv:1004.1063}]}.

\bibitem{Thiemann:1996aw}
T.~Thiemann, {Quantum spin dynamics (QSD)},
  \href{http://dx.doi.org/10.1088/0264-9381/15/4/011}{Class. Quant. Grav.
  {\bfseries 15}, 839 (1998)}, \href{http://arxiv.org/abs/gr-qc/9606089}{{
  [arXiv:gr-qc/9606089}]}.

\bibitem{Thiemann:1997rt}
T.~Thiemann, {Quantum spin dynamics (QSD): V. Quantum gravity as the natural
  regulator of matter quantum field theories},
  \href{http://dx.doi.org/10.1088/0264-9381/15/5/012}{Class. Quant. Grav.
  {\bfseries 15}, 1281 (1998)}, \href{http://arxiv.org/abs/gr-qc/9705019}{{
  [arXiv:gr-qc/9705019}]}.

\bibitem{Yang:2015zda}
J.~Yang and Y.~Ma, {New Hamiltonian constraint operator for loop quantum
  gravity}, \href{http://dx.doi.org/10.1016/j.physletb.2015.10.062}{Phys. Lett.
  B {\bfseries 751}, 343 (2015)}, \href{http://arxiv.org/abs/1507.00986}{{
  [arXiv:1507.00986}]}.

\bibitem{Alesci:2015wla}
E.~Alesci, M.~Assanioussi, J.~Lewandowski, and I.~M{\"a}kinen, {Hamiltonian
  operator for loop quantum gravity coupled to a scalar field},
  \href{http://dx.doi.org/10.1103/PhysRevD.91.124067}{Phys. Rev. D {\bfseries
  91}, 124067 (2015)}, \href{http://arxiv.org/abs/1504.02068}{{
  [arXiv:1504.02068}]}.

\bibitem{Alesci:2011ia}
E.~Alesci, T.~Thiemann, and A.~Zipfel, {Linking covariant and canonical LQG:
  New solutions to the Euclidean scalar constraint},
  \href{http://dx.doi.org/10.1103/PhysRevD.86.024017}{Phys. Rev. D {\bfseries
  86}, 024017 (2012)}, \href{http://arxiv.org/abs/1109.1290}{{
  [arXiv:1109.1290}]}.

\bibitem{Thiemann:2013lka}
T.~Thiemann and A.~Zipfel, {Linking covariant and canonical LQG II: Spin foam
  projector}, \href{http://dx.doi.org/10.1088/0264-9381/31/12/125008}{Class.
  Quant. Grav. {\bfseries 31}, 125008 (2014)},
  \href{http://arxiv.org/abs/1307.5885}{{ [arXiv:1307.5885}]}.

\bibitem{Zhang:2018wbc}
C.~Zhang, J.~Lewandowski, and Y.~Ma, {Towards the self-adjointness of a
  Hamiltonian operator in loop quantum gravity},
  \href{http://dx.doi.org/10.1103/PhysRevD.98.086014}{Phys. Rev. D {\bfseries
  98}, 086014 (2018)}, \href{http://arxiv.org/abs/1805.08644}{{
  [arXiv:1805.08644}]}.

\bibitem{Zhang:2019dgi}
C.~Zhang, J.~Lewandowski, H.~Li, and Y.~Ma, {Bouncing evolution in a model of
  loop quantum gravity},
  \href{http://dx.doi.org/10.1103/PhysRevD.99.124012}{Phys. Rev. D {\bfseries
  99}, 124012 (2019)}, \href{http://arxiv.org/abs/1904.07046}{{
  [arXiv:1904.07046}]}.

\bibitem{Soo:2005gw}
C.~Soo, {Further simplification of the super-Hamiltonian constraint of general
  relativity, and a reformulation of the Wheeler-Dewitt equation},
  \href{http://arxiv.org/abs/gr-qc/0512025}{{ arXiv:gr-qc/0512025}}.

\bibitem{Soo:2007hj}
C.~Soo, {Three-geometry and reformulation of the Wheeler-DeWitt equation},
  \href{http://dx.doi.org/10.1088/0264-9381/24/6/011}{Class. Quant. Grav.
  {\bfseries 24}, 1547 (2007)}, \href{http://arxiv.org/abs/gr-qc/0703074}{{
  [arXiv:gr-qc/0703074}]}.

\bibitem{Ashtekar:2003hd}
A.~Ashtekar, M.~Bojowald, and J.~Lewandowski, {Mathematical structure of loop
  quantum cosmology}, \href{http://dx.doi.org/10.4310/ATMP.2003.v7.n2.a2}{Adv.
  Theor. Math. Phys. {\bfseries 7}, 233 (2003)},
  \href{http://arxiv.org/abs/gr-qc/0304074}{{ [arXiv:gr-qc/0304074}]}.

\bibitem{Bojowald:2001xe}
M.~Bojowald, {Absence of singularity in loop quantum cosmology},
  \href{http://dx.doi.org/10.1103/PhysRevLett.86.5227}{Phys. Rev. Lett.
  {\bfseries 86}, 5227 (2001)}, \href{http://arxiv.org/abs/gr-qc/0102069}{{
  [arXiv:gr-qc/0102069}]}.

\bibitem{Ashtekar:2006rx}
A.~Ashtekar, T.~Pawlowski, and P.~Singh, {Quantum nature of the big bang},
  \href{http://dx.doi.org/10.1103/PhysRevLett.96.141301}{Phys. Rev. Lett.
  {\bfseries 96}, 141301 (2006)}, \href{http://arxiv.org/abs/gr-qc/0602086}{{
  [arXiv:gr-qc/0602086}]}.

\bibitem{Ashtekar:2006uz}
A.~Ashtekar, T.~Pawlowski, and P.~Singh, {Quantum nature of the big bang: An
  analytical and numerical investigation.},
  \href{http://dx.doi.org/10.1103/PhysRevD.73.124038}{Phys. Rev. D {\bfseries
  73}, 124038 (2006)}, \href{http://arxiv.org/abs/gr-qc/0604013}{{
  [arXiv:gr-qc/0604013}]}.

\bibitem{Ashtekar:2006wn}
A.~Ashtekar, T.~Pawlowski, and P.~Singh, {Quantum nature of the big bang:
  Improved dynamics}, \href{http://dx.doi.org/10.1103/PhysRevD.74.084003}{Phys.
  Rev. D {\bfseries 74}, 084003 (2006)},
  \href{http://arxiv.org/abs/gr-qc/0607039}{{ [arXiv:gr-qc/0607039}]}.

\bibitem{Ding:2008tq}
Y.~Ding, Y.~Ma, and J.~Yang, {Effective scenario of loop quantum cosmology},
  \href{http://dx.doi.org/10.1103/PhysRevLett.102.051301}{Phys. Rev. Lett.
  {\bfseries 102}, 051301 (2009)}, \href{http://arxiv.org/abs/0808.0990}{{
  [arXiv:0808.0990}]}.

\bibitem{Yang:2009fp}
J.~Yang, Y.~Ding, and Y.~Ma, {Alternative quantization of the Hamiltonian in
  loop quantum cosmology},
  \href{http://dx.doi.org/10.1016/j.physletb.2009.10.072}{Phys. Lett. B
  {\bfseries 682}, 1 (2009)}, \href{http://arxiv.org/abs/0904.4379}{{
  [arXiv:0904.4379}]}.

\bibitem{Assanioussi:2018hee}
M.~Assanioussi, A.~Dapor, K.~Liegener, and T.~Paw{\l}owski, {Emergent de Sitter
  epoch of the quantum cosmos from loop quantum cosmology},
  \href{http://dx.doi.org/10.1103/PhysRevLett.121.081303}{Phys. Rev. Lett.
  {\bfseries 121}, 081303 (2018)}, \href{http://arxiv.org/abs/1801.00768}{{
  [arXiv:1801.00768}]}.

\bibitem{Dapor:2017rwv}
A.~Dapor and K.~Liegener, {Cosmological effective Hamiltonian from full loop
  quantum gravity dynamics},
  \href{http://dx.doi.org/10.1016/j.physletb.2018.09.005}{Phys. Lett. B
  {\bfseries 785}, 506 (2018)}, \href{http://arxiv.org/abs/1706.09833}{{
  [arXiv:1706.09833}]}.

\bibitem{Li:2018opr}
B.-F. Li, P.~Singh, and A.~Wang, {Towards cosmological dynamics from loop
  quantum gravity}, \href{http://dx.doi.org/10.1103/PhysRevD.97.084029}{Phys.
  Rev. D {\bfseries 97}, 084029 (2018)},
  \href{http://arxiv.org/abs/1801.07313}{{ [arXiv:1801.07313}]}.

\bibitem{Taveras:2008ke}
V.~Taveras, {Corrections to the Friedmann equations from loop quantum gravity
  for a universe with a free scalar field},
  \href{http://dx.doi.org/10.1103/PhysRevD.78.064072}{Phys. Rev. D {\bfseries
  78}, 064072 (2008)}, \href{http://arxiv.org/abs/0807.3325}{{
  [arXiv:0807.3325}]}.

\bibitem{Ashtekar:2002sn}
A.~Ashtekar, S.~Fairhurst, and J.~L. Willis, {Quantum gravity, shadow states,
  and quantum mechanics},
  \href{http://dx.doi.org/10.1088/0264-9381/20/6/302}{Class. Quant. Grav.
  {\bfseries 20}, 1031 (2003)}, \href{http://arxiv.org/abs/gr-qc/0207106}{{
  [arXiv:gr-qc/0207106}]}.

\end{thebibliography}
\end{document}